\documentclass{PoS}
\usepackage{graphicx}
\usepackage{amsmath}
\usepackage{enumerate}

\DeclareMathOperator{\Tr}{Tr}
\DeclareMathOperator{\tr}{Tr}

\DeclareMathOperator{\U}{U}
\DeclareMathOperator{\SU}{SU}

\newcommand{\be}{\begin{equation}}
\newcommand{\ee}{\end{equation}}
\newcommand{\bea}{\begin{eqnarray}}
\newcommand{\eea}{\end{eqnarray}}

\def\Xint#1{\mathchoice
   {\XXint\displaystyle\textstyle{#1}}%
   {\XXint\textstyle\scriptstyle{#1}}%
   {\XXint\scriptstyle\scriptscriptstyle{#1}}%
   {\XXint\scriptscriptstyle\scriptscriptstyle{#1}}%
   \!\int}
\def\XXint#1#2#3{{\setbox0=\hbox{$#1{#2#3}{\int}$}
     \vcenter{\hbox{$#2#3$}}\kern-.5\wd0}}

\def\dashint{\Xint-}

\title{Single site model of large N gauge theories coupled to adjoint fermions}

\ShortTitle{Single site model of large N gauge theories coupled to adjoint fermions}

\author{Robert Lohmayer\thanks{Research supported in part
    by the NSF under grant numbers PHY-0854744 and PHY-1205396.}\\
        Department of Physics. Florida International University, Miami
        FL 33199, USA\\
        E-mail: \email{robert.lohmayer@gmx.net}}

\author{\speaker{Rajamani Narayanan}\thanks{Research supported in part
    by the NSF under grant numbers PHY-0854744 and PHY-1205396.}\\
        Department of Physics. Florida International University, Miami
        FL 33199, USA\\
        E-mail: \email{rajamani.narayanan@fiu.edu}}

\abstract{We consider a single site large N gauge theory coupled to
  adjoint fermions at weak coupling. We study the distribution of the eigenvalues of the link
  variables using a four-dimensional
  density function. We show that it is possible to recover the
  infinite-volume continuum limit for a certain range of fermion
  flavors if we use fermions with a bare mass of zero. 
}

\FullConference{31st International Symposium on Lattice Field Theory - LATTICE 2013\\
		July 29 - August 3, 2013\\
		Mainz, Germany}

\begin{document}

\section{Introduction}
Large N gauge theory with fermions in the adjoint representation are interesting in
many ways. It provides a connection to gravity and string
theories~\cite{Aharony:1999ti}.
One can use it to understand the transition from conformal to
confining field theories~\cite{Patella:2011kp,Hietanen:2008mr}.
There is a possibility to study continuum gauge theories using a
matrix model~\cite{Kovtun:2007py}. In addition, it is possible to
numerically investigate a theory with a real number of fermion
flavors~\cite{Hietanen:2009ex}.

The two main questions pertaining to the matrix model are:
\begin{enumerate}
\item What is the range of fermion flavors for which the single-site 
massless theory can be expected to reproduce the infinite-volume 
continuum theory?
\item Can we reproduce the infinite-volume continuum theory with
  massive fermions?
\end{enumerate} 
We will provide an answer to both these questions in the weak coupling
limit and it is a summary of the results presented
in~\cite{Lohmayer:2013spa}.

\section{Weak coupling analysis and the density function}

We refer the reader to~\cite{Lohmayer:2013spa} for the details of the
single site model. 
The total action depends on $d$ $\SU(N)$ matrices and the gauge transformation is
\begin{align}
U_\mu \to g U_\mu g^\dagger.
\end{align}
The action has an additional $\U^d(1)$ symmetry given by
\begin{align} 
U_\mu \to e^{i\alpha_\mu} U_\mu\label{znsymm}
\end{align} 
with $ 0 \leq \alpha_\mu < 2\pi$. Restricting $\alpha_\mu$ to
$\frac{2\pi k_\mu}{N}$ with integers $0 \le k_\mu < N$ keeps it in
$\SU(N)$; otherwise we have trivially extended the $\SU(N)$ theory to a
$\U(N)$ theory. 
Note that the eigenvalues of $U_\mu$ are gauge invariant.

In order to figure out if we can reproduce continuum infinite volume
results at the level of perturbation theory, we
follow~\cite{Bhanot:1982sh}
and set
\begin{align}
U_\mu = V_\mu D_\mu V^\dagger_\mu\,;\ \ \ \  
D_\mu^{jk} = e^{i\theta^j_\mu}\delta^{jk}.\label{upert}
\end{align}
We replace the integral over $U_\mu$ by an integral over $V_\mu$ and
$\theta_\mu^j$. We write $V_\mu=e^{ia_\mu}$ with $a_\mu^\dagger =
a_\mu$ and $a_\mu^{ii}=0$ for all $i$ 
and expand in powers of $a_\mu$ to compute observables in perturbation
theory.
We then have to show that the integral over $\theta_\mu^j$ is
dominated in the large N limit such that continuum infinite volume
perturbation theory is reproduced order by order.
We restrict ourselves to the lowest order in perturbation theory.

\begin{figure}[ht]
\centerline{
\includegraphics[width=140mm]{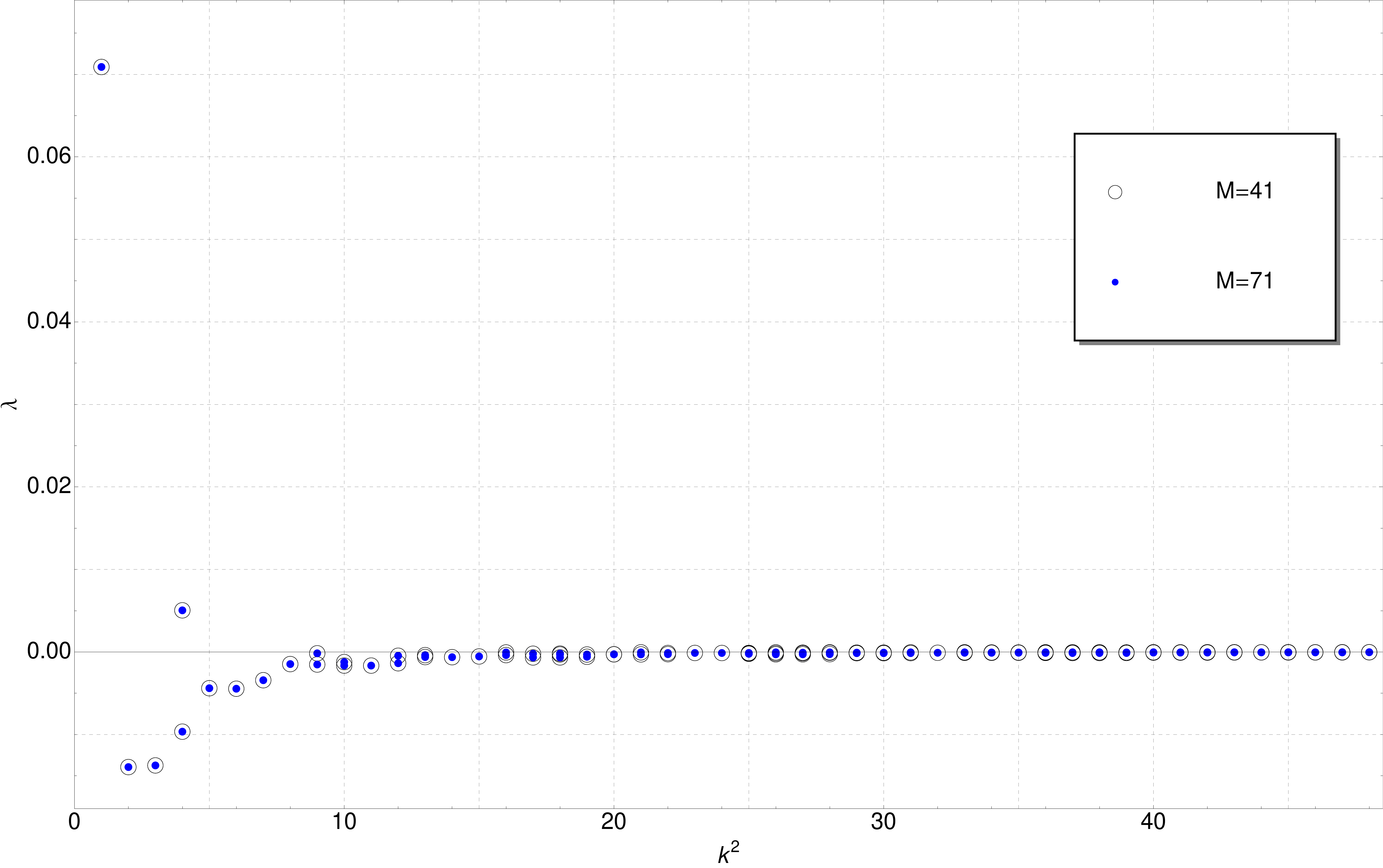}
}\caption{
Eigenvalues $\lambda_k = \lambda_k^{(g)} + f \lambda_k^{(o)}$ as a function of $k^2$ for the massless overlap Dirac operator with $f=1$ and
$m_w=-1$ obtained using numerical integration with $M^4$ equally
spaced points in the four-dimensional integration space.
} \label{fig1}
\end{figure}

As $N\to\infty$, we assume that we can define a joint distribution,
$\rho(\theta)$, in the following sense:
At any finite $N$, for a fixed choice of $\theta_\mu^j$, $j=1,\ldots,N$ and
$\mu=1,\ldots,4$, let
\begin{align}
\rho(\theta) = \frac{1}{N} \sum_j \prod_\mu
\delta(\theta_\mu-\theta_\mu^j)\,;\qquad\qquad \int \prod_\mu d\theta_\mu \rho(\theta) =1\,,\label{distdef}
\end{align}
where $\delta$ denotes the $2\pi$-periodized delta function normalized to $\int_{-\pi}^\pi d\theta \delta(\theta)=1$.
At the lowest order in perturbation theory (subscripts $g$ and $f$ stand for the
gauge and fermion contributions in the equations below),
\begin{align}
S^0_{g,f} [\rho] &= N^2 
\dashint d^4\theta d^4\phi\,
\rho(\theta) S_{g,f}(\theta-\phi) \rho(\phi)\,;
\cr
S_g(\theta) &= -\ln \hat p \,; \qquad\hat p =
\sum_\mu 4 \sin^2 \frac{\theta_\mu}{2}\,;\cr
S_f(\theta) &= 2 \ln \gamma_{w,o}(m_{w,o})\,; \cr
\gamma_w (m_w)&=\left(m_w+\frac{\hat p}{2} \right)^2 + \bar p\,  ;  \qquad
\bar p  =  \sum_\mu \sin^2 \theta_\mu\,;
\cr
\gamma_o(m_o,m_w) & = \frac{1+m_o^2}{2}+\frac{1-m_o^2}{2} \frac{m_w+\frac{\hat
    p}{2} }{\sqrt{\gamma_w(m_w)}}\, . \label{dentree}
\end{align}

We now assume that, as $N\to\infty$, the partition function will be
dominated by a single distribution $\rho(\theta)$, maximizing
$S^0[\rho]=S^0_g[\rho]+f S^0_f[\rho]$ for $f$ flavors of fermions.
We will only allow distributions that are non-negative everywhere with
the normalization condition in (\ref{distdef}). Furthermore, we assume
that the dominating distribution $\rho(\theta)$ is smooth and finite for all $\theta$ (in contrast to $\rho$ defined in \eqref{distdef} for angle configurations at finite $N$).
Clearly, $S^0$ in (\ref{dentree}) is invariant under $\rho(\theta) \to
\rho(\theta+\alpha)$ for any choice of $\alpha$, corresponding
to the invariance under  (\ref{znsymm}).

\begin{figure}[ht]
\centerline{
\includegraphics[width=140mm]{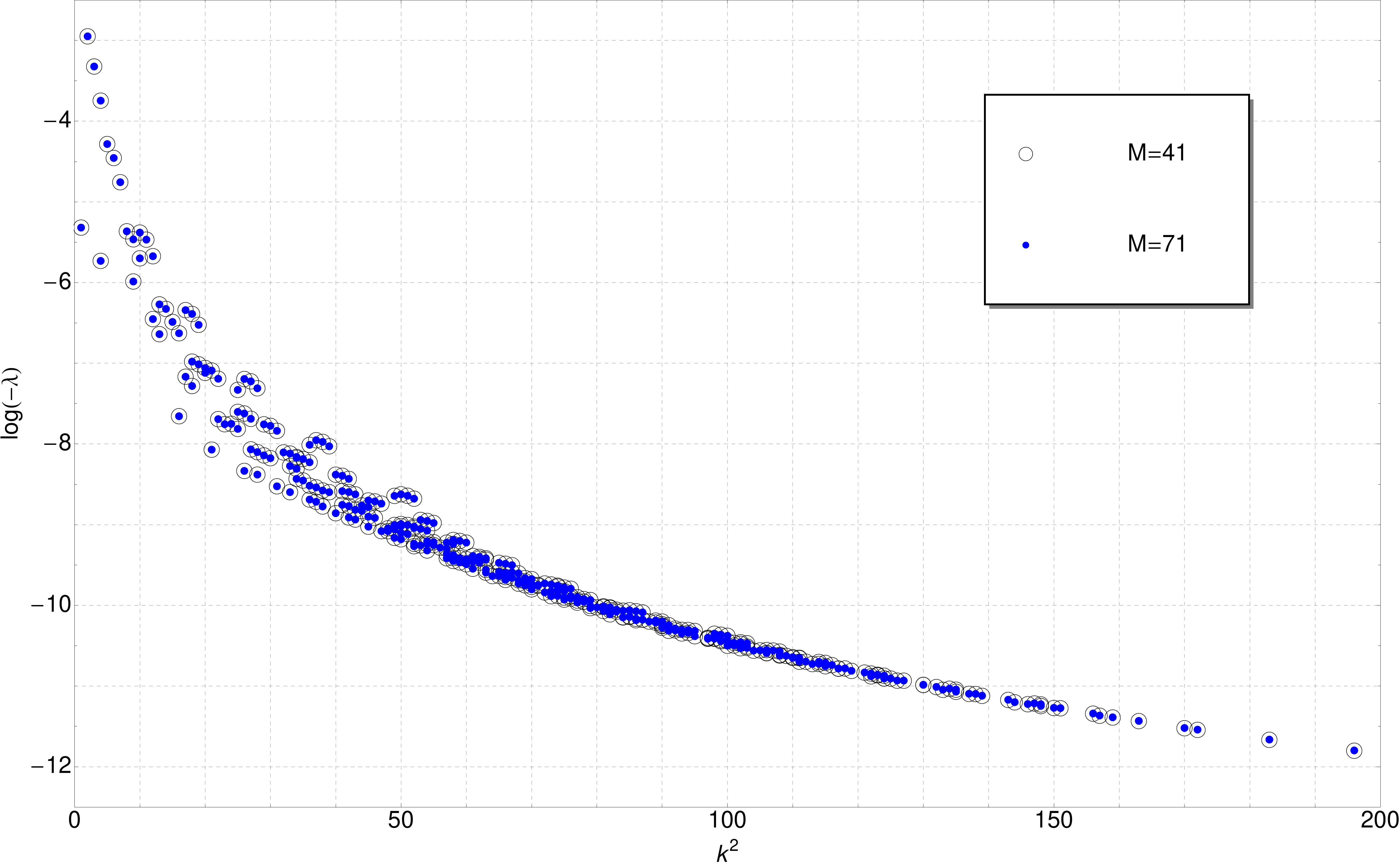}
}\caption{
Logarithm of the eigenvalues for the massless overlap Dirac operator with $f=2$ and
$m_w=-1$ obtained using numerical integration with $M^4$ equally
spaced points in the four-dimensional integration space.
} \label{fig2}
\end{figure}

Owing to the periodic and symmetric nature of $S_{g,f}(\theta)$, it follows that
\begin{align}
&\int_{-\pi}^\pi \prod_\nu \frac{d\phi_\nu}{2\pi}\, S_{g,f}(\theta-\phi)\, e^{i \sum_\mu k_\mu \phi_\mu}
= \lambda^{(g,f)}_{k}\, e^{i \sum_\mu k_\mu \theta_\mu}\,;\label{eigen}\\
\lambda^{(g,f)}_k = &\int_0^\pi \prod_\nu \frac{d\phi_\nu}{\pi}\,
S_{g,f}(\phi) \prod_\mu \cos(k_\mu\phi_\mu)\,. 
\label{eigen2}
\end{align}
Therefore, Fourier expanding
\begin{align}\label{eq:FourierExpansion}
\rho(\theta)= \frac1{(2\pi)^4} \sum_k c_k e^{i \sum_\mu k_\mu \theta_\mu} \qquad\textnormal{with}\qquad c_{-k}=c_k^\ast\,,\quad c_0=1
\end{align}
results in 
\begin{align}\label{eq:S-ck}
S^0_{g,f}=N^2 \sum_k c_kc_k^\ast \lambda_k^{(g,f)}.
\end{align}

If all the eigenvalues, 
\begin{align}
\lambda_k=\lambda^{(g)}_k + f \lambda^{(f)}_k
\end{align} 
for $k\ne 0$ are smaller than zero,
the constant mode, $\rho(\theta) = \frac{1}{(2\pi)^4}$, will dominate
in the large-$N$ limit (i.e., $c_k\to 0$ for $k\neq0$) and the single-site model will be in the
correct continuum phase and possibly reproduce the infinite-volume
continuum theory. 

\begin{figure}[ht]
\centerline{
\includegraphics[width=140mm]{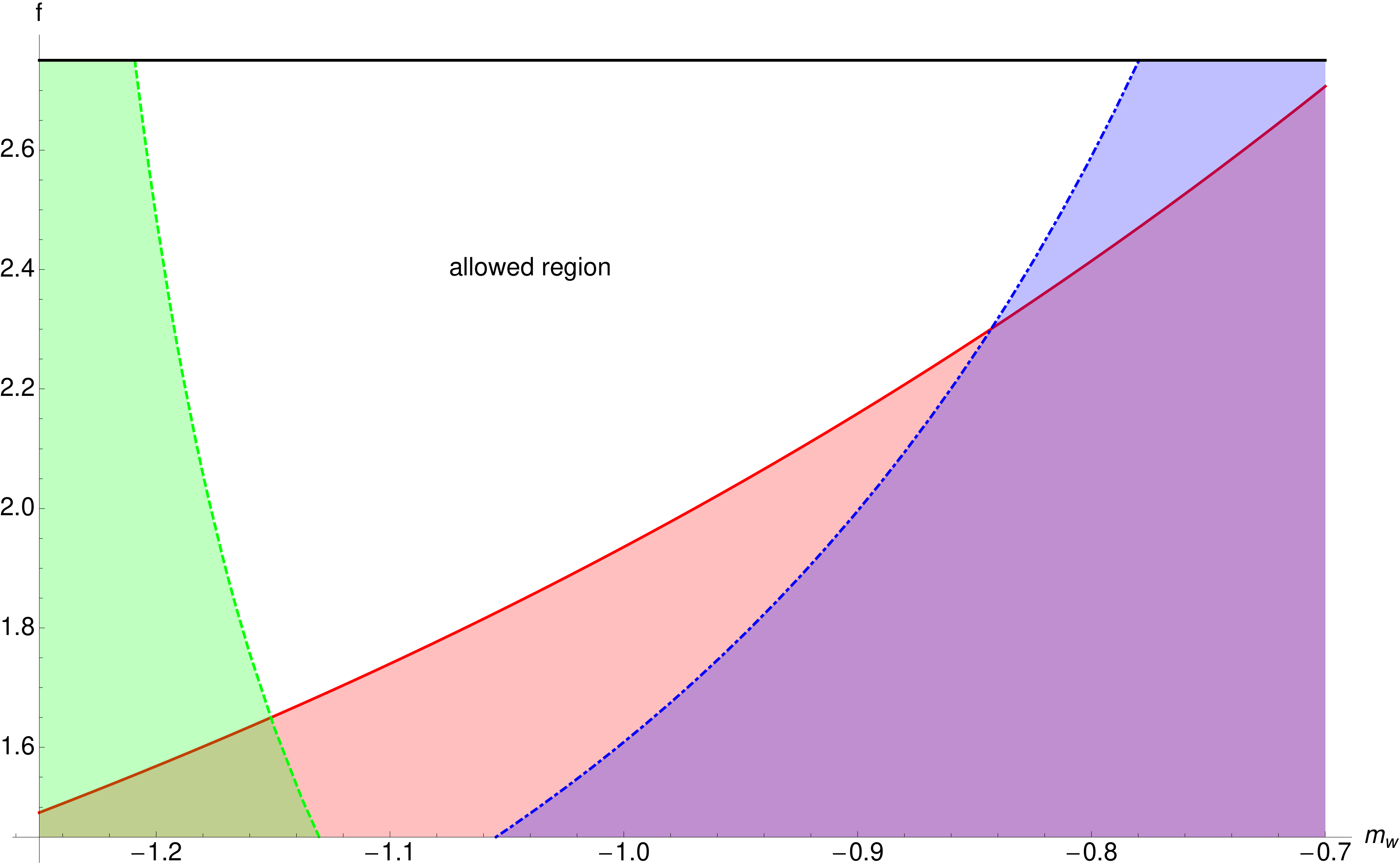}
}\caption{
Boundary of the allowed region in the $(m_w,f)$-plane for massless overlap fermions ($m_o=0$). The different lines show $- \lambda^{(g)}_k/\lambda^{(o)}_k$ for $k=(2,2,2,2)$ (green, dashed), $k=(1,0,0,0)$ (red, solid), and $k=(2,0,0,0)$ (blue, dot-dashed). The intersection points are at $(m_w,f)\approx(-0.843,2.30)$ and $(m_w,f)\approx(-1.15,1.65)$.  
} \label{fig:allowed}
\end{figure}

If some of the eigenvalues are larger than zero, then the action $S^0$
in (\ref{dentree}) will not be maximized by $\rho(\theta) =
\frac{1}{(2\pi)^4}$ and some $c_k$ ($k\neq0$) will be non-zero. Since
the action in \eqref{eq:S-ck} is quadratic, the maximum will be
obtained at the boundary of the domain of allowed values for the
$c_k$'s, which is determined by the condition $\rho(\theta)\geq0$ for
all $\theta$. Therefore, $S[\rho]$ will be maximized by a
$\rho(\theta)$ which is zero at least at one point in
the four-dimensional Brillouin zone. Due to the shift-invariance, there will then be a
class of densities, related by $\rho(\theta)\to\rho(\theta+\alpha)$
with arbitrary $\alpha$, having identical maximum action resulting in
a spontaneous breaking of the $\U^d(1)$ symmetry in (\ref{znsymm}).

\begin{figure}[ht]
\centerline{
\includegraphics[width=140mm]{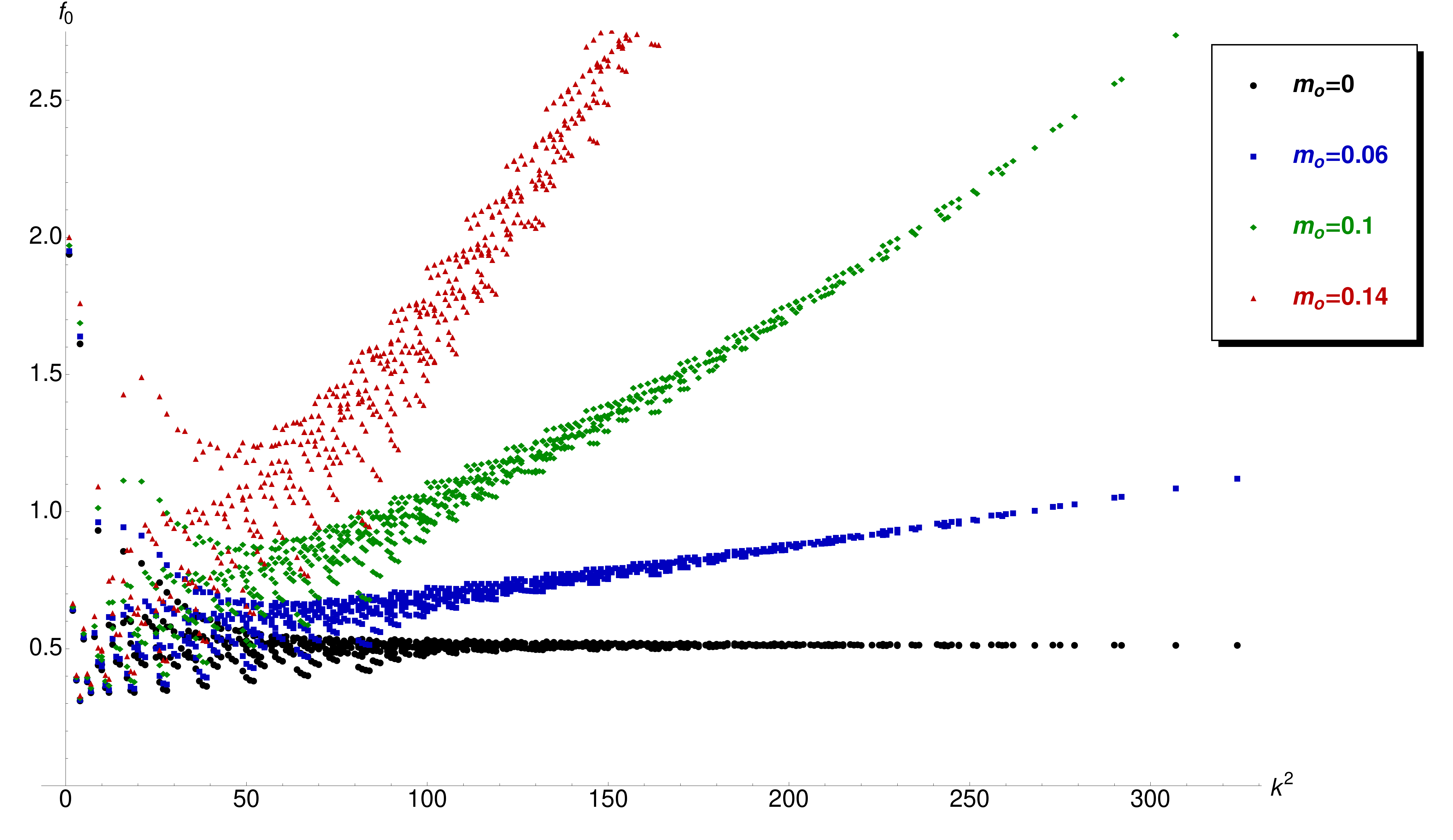}
}\caption{
Plots of $f_0\equiv - \lambda^{(g)}_k/\lambda^{(o)}_k$ (for $k_\mu\leq 9$) at $m_w=-1$ and different choices for $m_o$. For $m_o=0$, $f_0\to0.5$ as $k^2\to\infty$; for all $m_o>0$, $f_0\to\infty$ as $k^2\to\infty$. The boundary of the allowed region is determined by $\max_k f_0(k)$.
} \label{fig:overlap-f-vs-mo}
\end{figure}
\section{Investigation of the allowed regions}

We only consider the case of overlap fermions and refer the reader to
\cite{Lohmayer:2013spa} for the case of Wilson fermions.
A sample plot is shown in Fig.~\ref{fig1} where we have computed the
eigenvalues $\lambda_k=\lambda_k^{(g)}+f\lambda_k^{(o)}$ for massless overlap fermions with $f=1$ and $m_w=-1$.
The results are obtained with $M^4$ equally spaced points in the 
four-dimensional integration space and we used $M=41$ and $M=71$ to
show that we have reached the limit of the continuum integral.
Since two eigenvalues are positive, $(f=1,m_o=0,m_w=-1)$ is not a point in the allowed region for overlap fermions.

As a second example, we set $f=2$, keeping $m_o=0$ and $m_w=-1$. In this case, we find all eigenvalues $\lambda_k$ to be negative, making this a point inside the allowed region. In Fig.~\ref{fig2}, we have plotted $\ln(-\lambda_k)$ as a function of $k^2$ to show that even in the log-scale we have a good estimate for the continuum integral.   

Numerically, we find that $\lambda_k^{(g)}>0$ for all $k$, which means that a point $(f,m_o,m_w)$ will be inside the allowed region (defined by $\lambda_k=\lambda_k^{(g)}+f\lambda_k^{(o)}<0$ for all $k$) iff
\begin{enumerate}[(i)]
\item  $\lambda_k^{(o)}(m_o,m_w)<0$ for all $k$,
\item $f>\max_k \left\{ - \lambda^{(g)}_k/\lambda^{(o)}_k \right\}$.
\end{enumerate}      
For massless overlap fermions,
$- \lambda^{(g)}_k/\lambda^{(o)}_k\to \frac 12$ as $k\to\infty$ for
all $m_w<0$. Therefore, for $m_o=0$, the allowed region in the $(m_w,f)$-plane is determined by eigenvalues $\lambda_k$ with $k$ being small.
A plot of the boundary of the allowed region in the $(m_w,f)$-plane
for $m_o=0$ is shown in Fig.~\ref{fig:allowed}. 

For $m_o>0$, $\lambda_k^{(o)}(m_o)/\lambda_k^{(o)}(m_o=0)\to 0$ as
$k\to\infty$. 
Together with our results for the massless case, this immediately
implies that $- \lambda^{(g)}_k/\lambda^{(o)}_k\to\infty$ as
$k\to\infty$ (see Fig.~\ref{fig:overlap-f-vs-mo} for numerical
results). 
Therefore, it is necessary to keep $m_0=0$ in the weak-coupling limit.

\section {Conclusions}

Previous numerical work considered quantities like $\Tr U_\mu$, $\Tr
U_\mu U_\nu$, $\tr U_\mu U_\mu^\dagger$ and a few others. These
correspond to a select set of values of $k_\mu$ in the weak coupling limit.
We have
shown that this can lead to incorrect conclusions about the validity of
the single site model.
Since some coefficients with small $k$ could be accidentally small, even
looking at $k_μ=(1,-1,0,0)$
 might not be sufficient to check if the single site model can reproduce the infinite volume continuum theory.
Much of the previous numerical work has been done at finite values of
the lattice coupling. Even if there is some evidence for an infinite
volume limit at finite lattice coupling, the results here show that
one cannot take the weak coupling limit. This also applies to
numerical work done with massive fermions.

\end{document}